\documentclass[iop]{emulateapj}

\usepackage{lscape}

\slugcomment{accepted to ApJ}

\begin{document}

\title{The substellar population of $\sigma$ Orionis: A deep wide survey}

\author{B\'ejar, V. J. S.}
\affil{Instituto de Astrof\'{\i}sica de Canarias, E-38205 La Laguna, 
Tenerife, Spain}
\affil{Departamento de Astrof\'{\i}sica, Universidad de La Laguna, 38205. La Laguna,
Tenerife, Spain}
\email{vbejar@iac.es}

\author{Zapatero Osorio, M. R.}
\affil{Centro de Astrobiolog\'{\i}a (INTA-CSIC), Crta. Ajalvir km 4, E-28850 Torrej\'on 
de Ardoz, Madrid, Spain}
\email{mosorio@cab.inta-csic.es}

\author{Rebolo, R.} 
\affil{Instituto de Astrof\'{\i}sica de Canarias, E-38205 La Laguna, 
Tenerife, Spain}
\affil{Departamento de Astrof\'{\i}sica, Universidad de La Laguna, 38205. La Laguna,
Tenerife, Spain}
\affil{Consejo Superior de Investigaciones Cient\'{\i}ficas, CSIC, Spain}
\email{rrl@iac.es}

\author{Caballero, J. A.} 
\affil{Centro de Astrobiolog\'{\i}a (INTA-CSIC), ESAC campus, PO Box 78, E-28691 Villanueva de la Ca\~{n}ada, Madrid, Spain}
\email{caballero@cab.inta-csic.es}

\author{Barrado, D.}
\affil{Centro de Astrobiolog\'{\i}a (INTA-CSIC), ESAC campus, PO Box 78, E-28691 Villanueva de la Ca\~{n}ada, Madrid, Spain}
\affil{Calar Alto Observatory, Centro Astron\'omico Hispano Alem\'an, 
C/Jes\'us Durb\'an Rem\'on, 2-2, E-04004 Almer\'{\i}a, Spain }
\email{barrado@cab.inta-csic.es}

\author{Mart\'{\i}n, E. L.}
\affil{Centro de Astrobiolog\'{\i}a (INTA-CSIC), Crta. Ajalvir km 4, E-28850 Torrej\'on 
de Ardoz, Madrid, Spain}
\email{ege@cab.inta-csic.es}

\author{Mundt, R.}
\affil{Max-Planck-Institut f\"ur Astronomie, K\"onigstuhl 17, D-69117
Heidelberg, Germany}
\email{mundt@mpia.de}

\and

\author{Bailer-Jones, C. A. L.}
\affil{Max-Planck-Institut f\"ur Astronomie, K\"onigstuhl 17, D-69117
Heidelberg, Germany}
\email{calj@mpia.de}

\begin{abstract}
    
We present a deep $I$,$Z$ photometric survey covering a total area 
of 1.12 deg$^2$ of the $\sigma$ Orionis cluster and 
reaching completeness magnitudes of $I$=22 and $Z$=21.5\,mag. 
From $I$, $I-Z$ color--magnitude diagrams we have selected  153 candidates that fit 
the previously known sequence of the cluster. They have magnitudes in 
the range $I$=16--23\,mag, which corresponds to a mass interval from 0.1 down to 0.008 
M$_{\odot}$ at the most probable age of $\sigma$ Orionis (2--4 Myr).
Using $J$-band photometry, we find that 124  of the 151 candidates within the completeness of 
the optical survey (82\,\%) follow the previously known infrared photometric sequence of the cluster and are probably
members. We have studied the spatial distribution of the very low mass stars and brown dwarf population of the cluster and found that 
there are objects located at distances greater than 30 arcmin 
to the north and west of $\sigma$ Orionis that probably belong to different populations of the Orion's Belt. 
For the 102 bona fide $\sigma$ Orionis cluster member candidates, we find that 
the radial surface density can be represented by a decreasing exponential  function ($\sigma = \sigma_{0} e^{-r/r_{0}}$) with a central density of 
$\sigma_{0}$=0.23 $\pm$ 0.03 object/arcmin$^2$ and a characteristic radius of $r_{0}$=9.5 $\pm$ 0.7 arcmin. 
From a statistical comparison with Monte Carlo simulations, 
we  conclude that the spatial distribution of the objects located at the same distance from the center of the cluster 
is compatible with a Poissonian distribution and, hence, that very low mass stars and brown dwarfs are not mainly forming aggregations or sub-clustering. 
Using near-infrared $JHK$-band data from 2MASS and UKIDSS and mid-infrared data from IRAC/{\it Spitzer}, we find that 
about 5--9\,\% of the brown dwarf candidates in the $\sigma$ Orionis cluster have $K$-band excesses and 
31$\pm$7\,\% of them show mid-infrared excesses at wavelengths longer than 5.8\,$\mu$\,m. These are probably related to the presence of disks, most of which are 
``transition disks''. 
 We have also calculated the initial mass spectrum ($dN/dm$) of $\sigma$ Orionis from very low mass stars ($\sim$\,0.10 M$_{\odot}$)  
to the deuterium-burning mass limit (0.012--0.013 M$_{\odot}$), i.e., complete in  the entire brown dwarf regime. This mass spectrum is a rising  
function toward lower masses and can be represented by a power-law distribution 
($dN/dm$ $\propto$ $m^{-\alpha}$) with an exponent $\alpha$ of 0.7$\pm$ 0.3 for an age of 3 Myr.

\end{abstract}


\section{Introduction}

Studies of substellar populations in clusters have the advantage over those in the 
field that the objects have a common age, metallicity and origin and occupy a well defined region 
in the sky. These parameters are specially important for the study of  the 
mass function in the substellar regime, where the luminosity function depends drastically on 
the age. 

The very young $\sigma$ Orionis cluster, located around the star of the same name, 
has been known since the early studies of \cite{garr67} and \cite{lynga81,lynga83}. 
The $\sigma$ Orionis multiple star, whose brightest component is an O9.5V, 
belongs to the OB1b association in the Orion complex popularly known as  Orion's Belt. 
{\it R\"ontgensatellit (ROSAT)} satellite observations of this association led to the discovery of a very young 
stellar population around   $\sigma$ Orionis \citep{walter97,wolk00}. 
Previous photometric searches in the cluster (\citealt{bejar99}, hereafter BZOR; \citealt{bejar01}, hereafter BMZO; 
\citealt{bejar04a}; \citealt{sherry04}; \citealt{kenyon05}; \citealt{cab07a}; \citealt{cab08a}; \citealt{lodieu09}) 
have found a large stellar and substellar population. 
Follow-up spectroscopic studies have allowed us to characterize the spectral sequence of substellar members between M6 and T5.5 (BZOR; 
\citealt{osorio99a,osorio00,osorio02a,osorio02b}; \citealt{barrado01,barrado02a,barrado03}; 
\citealt{martin01}; \citealt{kenyon05}; \citealt{burn05}). 
Studies of the depletion of lithium in the atmosphere of K6--M8.5 spectral type low mass members of the cluster 
impose an upper age limit  of 8 Myr and suggest a most likely age for the cluster 
in the interval 2--4 Myr \citep{osorio02b}. This is in good agreement with previous age determinations 
based on more massive stars in the Orion association \citep{blaauw64,blaauw91,warren78,brown94}. 
Other studies, based on isochrone fits 
to photometric sequences, have determined similar ages of the cluster (see \citealt{wolk00}; BZOR; 
\citealt{oliveira02, oliveira04}; \citealt{sherry04}).  
The {\it Hipparcos} satellite provides a distance of 352$^{+166}_{-85}$\,pc \citep{perryman97} 
for the central star $\sigma$ Orionis\,AB. This distance agrees with previous works on  
the Orion OB1b association, which determine a distance between 360 and 500\,pc \citep{blaauw64,blaauw91,warren78,brown94}  
and with recent estimates based on main sequence fitting and 
the dynamical parallax of $\sigma$ Ori\,AB, which give distances of 444 $\pm$ 22\,pc, 400\,pc, and 334$^{+25}_{-22}$\,pc,\footnote{The real distance 
becomes $\sim$ 385\,pc if $\sigma$ Ori\,AB is a triple system \citep{cab08c}, as recently claimed by \cite{simon11}} respectively \citep{sherry08,mayne08,cab08c}. 
This star is affected by a low extinction of $E(B-V)$~=~0.05\,mag \citep{lee68}, and 
the associated cluster also exhibits very little reddening, with a typical visual extinction 
$A_{V}$ $<$ 1\,mag (see BMZO; \citealt{oliveira02}; \citealt{bejar04a}). 
In addition, the average metallicity of the  $\sigma$ Orionis cluster is determined to be [Fe/H]=--0.02$\pm$0.09$\pm$0.13 (random and systematic errors),
which is consistent with solar values \citep{jonay08}.   

The $\sigma$ Orionis star cluster is one of the best sites in which to define the substellar initial mass function 
because of its high number of the cluster members, which leads to good statistics in a relatively small area and 
a knowledge of the cluster sequence for a wide range of masses from 25 $M_{\odot}$ to 0.003 $M_{\odot}$;  
its low extinction, which allows us to assign directly masses from magnitudes in comparison with theoretical 
isochrones; and the absence of differential reddening, which would otherwise obscure some of its members. In spite of these 
unique characteristics, $\sigma$ Orionis has a very young age, which means that no dynamical evolution is expected, 
and that the mass function is very close to the initial mass function. Several studies  have dealt with the 
cluster mass function both in the stellar and substellar domain (BMZO; \citealt{gon06,cab07,cab07a,lodieu09,bihain09}). 
In addition, a lot of effort has been expended on this site to investigate the formation process of substellar objects, and 
in particular, the study of accretion disks and/or outflows \citep{barrado02a,barrado03,osorio02a,osorio02b,muzerolle03,kenyon05,cab06}, 
and the existence of infrared excesses 
possibly related with the presence of disks, both in the near-infrared \citep{oliveira02,barrado03,bejar04a}  
and in the mid-infrared \citep{jaya03,oliveira04,oliveira06,hernandez07,cab07a,osorio07,scholz08,luhman08}.

In this paper we present a deep $IZ(J)$ survey covering an area of 1.12  deg$^2$ around 
the star $\sigma$ Orionis. We aim to detect and characterize very low-mass stars and substellar objects with completeness in the 
whole brown dwarf domain in a significant area of the cluster. Details of the observations are indicated in Section 2. In Section 3 we explain the 
criteria for selecting member candidates. Section 4 is devoted to study the spatial distribution of candidates in the cluster. 
In section 5 we study their infrared excesses using the available Two Micron All Sky Survey (2MASS), 
UKIRT Deep Infrared Sky Survey (UKIDSS) and {\em Spitzer} photometry, and in 
Section 6 we estimate the substellar mass spectrum of the cluster. Conclusions are given in Section 7. 
 Part of the data on which the present paper is based were used in previous surveys by \cite{osorio00} and BMZO, 
covering an area of 847\,arcmin$^2$. Preliminary results of the cluster substellar spatial distribution were presented by \cite{bejar04b}.

\section{Observations}

\subsection{Optical Photometry}

\begin{figure}[]
\epsscale{0.8}
\vspace{1.75in}
\plotone{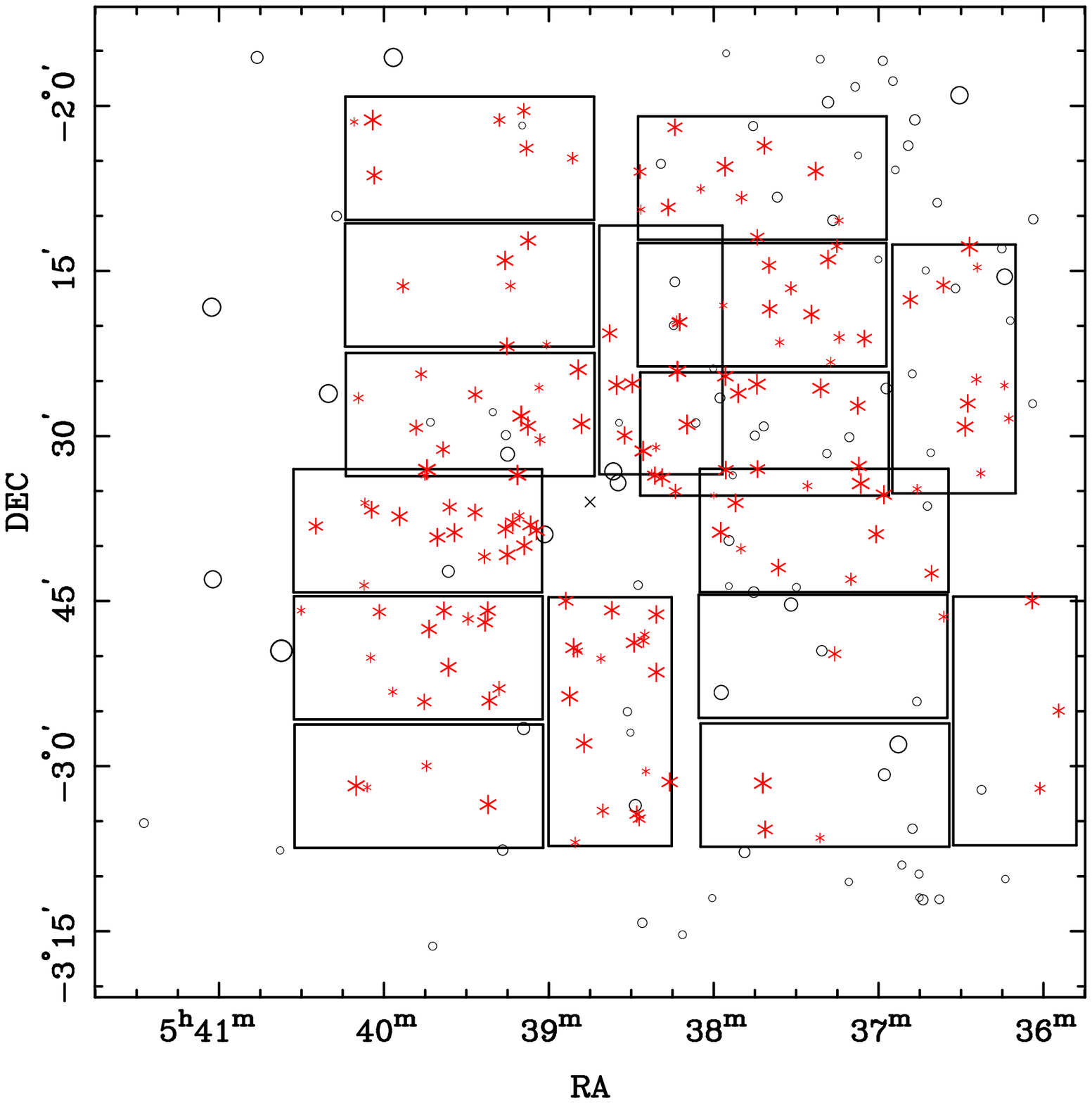}
\caption[]{\label{fig1} Representation of present survey in the $\sigma$ Orionis cluster
Area covered by the four detectors of each pointing are shown in squares. 
Field stars brighter than 12\,mag are indicated by open circles. The 153 selected optical candidates 
are denoted by stars and red colors (in the electronic version).
The size of the symbols is inversely proportional to the
$I$ magnitude.
The multiple star $\sigma$ Orionis is labeled with a cross indicating central
coordinates.}
\end{figure}

We obtained $IZ$ images with the Wide Field Camera (WFC) mounted on the Cassegrain focus of the Isaac Newton Telescope (INT) on 1998 November 12 and 13. 
The camera consists of a mosaic of four 2\,k$\times$4\,k Loral CCD detectors, providing a pixel projection 
of 0.33 and covering an effective area of 1012 arcmin$^2$ in each exposure. We
observed four different fields with tiny overlapping between neighboring pointings, 
covering a total area of 1.12 deg$^2$. The central coordinates 
of these pointings are indicated in Table \ref{tab1}. A representation of the survey can be seen in Figure \ref{fig1}. 
We performed three individual exposures of 1200\,s in each pointing, resulting in a total 
exposure time of 1 hour in each field and filter. 

Raw frames  were reduced within the IRAF\footnote{IRAF is distributed by National Optical Astronomy
Observatories, which is operated by the Association of Universities for
Research in Astronomy, Inc., under contract with the National Science
Foundation.} environment, using the {\tt CCDRED} package.
Images were bias subtracted, trimmed and flat-field corrected. We suitably combined 
our own long exposure scientific images  to obtain flat-fields. These flat images,
usually called superflats, are very useful for correcting fringing patterns not present 
in sky or dome flats. The photometric analysis was performed using routines within {\tt DAOPHOT}, which
include the selection of objects with stellar PSF using the {\tt DAOFIND} task (extended objects were mostly avoided) 
and aperture and PSF photometry. The average seeing on both nights varied from 1.0 to 
1.2 arcsec. Nights were not photometric and instrumental 
magnitudes were transformed into the real magnitudes in the Cousins $I$ system 
using observations of common stars obtained with the same instrumentation on a 
photometric night on 2003
January 8. This night was calibrated using photometric standard stars 
from Landolt \citep{landolt92}, observed throughout the night and in each of the four detectors. 
We found a difference in the zero points of the detectors, and hence, we calibrated  each of them independently. 
 Basically, the detector 4 (the one in the center) has systematically a zero point $\sim$\,0.4\,mag fainter, while the rest of them are similar within 0.1\,mag. 
The calibration of these data in BMZO was done assuming that the sensitivity of all the detectors were similar 
and this explains why the $I$-band photometry of 
some of the objects presented here is different from that presented in BMZO and \cite{osorio00}. 
Instrumental magnitudes of $Z$ filter were pseudo-transformed into apparent magnitudes assuming that the distribution
of the number of stars per interval of magnitude $I-Z$ is similar to that of the
Pleiades cluster \citep{osorio99b} and has a maximum around $I-Z
\sim$ 0.4\,mag. The absolute calibration of this filter is not strictly necessary for the
selection of our candidates, since this task is carried 
out in relative terms: for a given $I$-band magnitude, candidates must have $I-Z$ colors redder 
than field sources and overlap and extrapolate the expected photometric sequence of the cluster defined by known members. 
We always refer to the calibrated $I$ band to estimate masses for our objects. The survey completeness magnitudes are $I$=22.0, $Z$=21.5\,mag and the
limiting magnitudes are $I$=23.8, $Z$=23.0\,mag. We adopted as the completeness magnitude the value at which the histogram 
of detections as a function of magnitude reaches a maximum ($\sim$ 10-$\sigma$ detection),  
and as limiting magnitude the value at which 50\,\%~of the objects at the maximum of the 
histogram are detected ($\sim$ 3-$\sigma$ detection limit).

Table \ref{tab2} contains the optical photometry and coordinates of selected objects (see section \ref{sec3}). 
The error bars account for both the instrumental magnitude 
errors and the uncertainties in the photometric calibrations, which are typically 0.03-0.04\,mag. 
Astrometry was derived from 
the UKIDSS Galactic Cluster Survey (GSC) catalogue for those objects present in the Data Release 6 
(a correlation radius of 5 arcsec were used to cross-match the list of targets). 
The astrometry of fainter candidates not detected in UKIDSS was obtained from the plate solution of each detector derived using the UKIDSS
astrometry of bright objects in common and the {\tt CCMAP} routine. Typical root-mean-square of 0.1--0.4 arcsec were found in the astrometric solution 
for different detectors.

\subsection{Infrared Photometry} 

We obtained $J$-band photometry with the CAIN infrared camera  on the Telescopio 
Carlos S\'anchez (TCS) at Observatorio 
del Teide, on 1998 September 18, 1999 January 23, 24, February 24, August 22, 23, 24, November 26, 27, December 28, 29, 
and 2000 January 27, and February 11, 12, and with the MAGIC instrument mounted on the 2.2 m telescope at the Calar Alto 
Observatory on 1998 December 6. 
The CAIN camera consists of a 256$\times$256 pixel NICMOS3 infrared array, providing a pixel projection 
of 1.00 arcsec and covering a total area of 4.3$\times$4.3 arcmin$^2$ in each exposure. 
The MAGIC instrument has also a 256$\times$256 pixel NICMOS3 infrared array, providing a pixel projection 
of 0.64 arcsec and covering an area of 2.7$\times$2.7 arcmin$^2$. 
Exposure times ranged from 60 to 1000\,s (CAIN) and from 405 to 900\,s (MAGIC). 
Average seeing varied from 1.5 to 4.0 arcsec during the TCS observations and from 1.0 to 2.0 arcsec during the Calar Alto run. 
We also obtained $J$-band photometry for the complete area of the
BZOR survey on the 3.5 m Calar Alto telescope in 1998 October, using the Omega-Prime instrument, which was cross-matched  
with our optical photometry. See \cite{osorio00} and BMZO for more details 
about these $J$-band data. 
The raw CAIN and MAGIC data were processed within the IRAF environment, including sky subtraction and flat field correction. 
Final individual images were properly aligned and combined. Aperture photometry was 
performed using {\tt DAOPHOT} routines. Instrumental magnitudes were transformed into apparent magnitudes in the UKIRT system using 
several photometric field standards obtained for each night \citep{hunt98}. 
For some of our objects observed in non-photometric condidtions, the CAIN photometry was calibrated using 
the 2MASS photometry of objects in common in the same field of view. For a few objects for which the CAIN photometry has 
large uncertainties ($>$0.2\,mag) or there are no available data, we have adopted the photometry from UKIDSS (described below). 
All the available $J$-band data for our candidates are provided in Table \ref{tab2}. 
Error bars account for the instrumental magnitude errors and the uncertainties in the photometric calibrations, 
which are typically 0.02--0.13\,mag for the CAIN and MAGIC data and 0.05\,mag for Omega-Prime.

\subsubsection{2MASS and UKIDSS Near-Infrared Photometry} 
In addition to the TCS and 2.2m\,CA data, we have used the available 
$JHK_{s}$ photometry from the 2MASS All Sky Catalog of point sources \citep{cutri06} 
and $ZYJHK_{s}$ photometry from the UKIDSS GCS. 
The UKIDSS project is defined in \cite{lawrence07}. UKIDSS uses the UKIRT Wide Field Camera 
(WFCAM; \citealt{casali07}) and a photometric system described in \cite{hewett06}. 
The pipeline processing and science archive are described in 
J. Irwin, et al. (in preparation) and \cite{hambly08}. We have used data from the 6th data release (DR6plus). 
A radius of 5 arcsec was used to cross-match our list of candidates 
with corresponding catalogs. The $JHK_{s}$-band data for 97 objects were obtained from 
the 2MASS catalogue and $ZYJHK_{s}$ 
data for a total of 126 sources  were obtained from UKIDSS. Another 10 additional objects have available 
photometry in the $YJHK_{s}$ bands and another 10 have information in at least one of the UKIDSS GCS filters. 
For more details about the astrometric and photometric analysis of 2MASS data, see the Explanatory Supplement to the 2MASS All Sky Data Release 
\citep{cutri06}. For more details about the UKIDSS GCS and the astrometric and photometric quality of the data 
in the $\sigma$ Orionis cluster see previous work by \cite{lodieu09} and references therein. 
Completeness magnitudes of UKIDSS GCS in this cluster is $Z$=20.2, $Y$=20.0, $J$=19.0, $H$=18.4 and $K$=18.0\,mag, 
as estimated 
by \cite{lodieu09}. Our present survey is about 1.5\,mag deeper in the $Z$-band than UKIDSS GCS. This explains why there are 30 of our selected objects that have no 
UKIDSS counterpart in this band and 9 have no counterpart in any band. 
Two of these objects have an $I$-band magnitude fainter than 
the completeness of our survey, and four of them do not have a very red $I-J$ color. The other three are located 
at northern declinations (lower than -2$^{o}$01$^{'}$30$^{''}$) and are outside the 
UKIDSS GCS area in the $\sigma$ Orionis cluster. 
Nevertheless, most of the good cluster member candidates within the completeness of our survey have available photometry 
at least in the $JHK$-bands of the UKIDSS GCS catalogue. 
The available 2MASS photometry of selected objects is indicated in Table \ref{tab2}, while UKIDSS photometry is 
indicated in Table \ref{tab3}. 

To check the accuracy of our photometry and study possible variability among the selected candidates, 
we have compared the new $J$-band data presented in this paper with the photometry provided by the 2MASS and UKIDSS catalogs. 
Table \ref{tab4} summarizes the comparison between the CAIN and 2MASS photometry for the objects that have been
independently calibrated and between the CAIN and UKIDSS photometry.  In addition, we also include  
the difference between the $J$-band data from Omega-Prime and 2MASS and UKIDSS. In this table, the average differences, errors, 
standard deviation of the mean and the number 
of objects are indicated. 
In summary, the $J$-band photometry presented in this paper is consistent with the photometry of the 2MASS and UKIDSS 
catalogs. Small offsets can be explained by the differences in filter systems and the intrinsic variability of
 some of the targets.

\subsubsection{IRAC/Spitzer mid-Infrared Photometry} 

We have also used available public data from the Infrared Array Camera (IRAC, \citealt{fazio04}) on 
the {\em Spitzer Space Telescope}, belonging to the {\em Spitzer} Guarateed Time Observation program \#\,37 
(PI: G. Fazio). Processed images were downloaded using Leopard software. More details about these data can be found 
in \cite{hernandez07}. Aperture and PSF photometry were performed as indicated in \cite{osorio07}. 
A comparison of this photometry and that presented by \cite{luhman08} using the same 
data can be found in \cite{bihain09}. A radius of 5 arcsec was used to cross-match our list of objects  
with the IRAC/{\em Spitzer} data. For 98, 89, 83 and 71 sources there is 3.6, 4.5, 5.8 and 8.0\,$\mu$\,m 
photometry, respectively, while a total of 103 objects have photometry in at least one filter. Many of our
 candidates are too faint to 
be automatically detected in the IRAC/{\em Spitzer} images, especially in 5.8 and 8.0\,$\mu$\,m, but 
a large number of them are 
also outside the surveyed area in the mid-infrared, which 
 is also slightly different for the [3.6]/[5.4] and [4.5]/[8.0] pairs of bands. 
The available IRAC/{\em Spitzer} photometry of selected objects is given in Table \ref{tab3}.

\section{\label{sec3} Selection of cluster member candidates}

\begin{figure}[]
\epsscale{0.8}
\vspace{1.75in}
\plotone{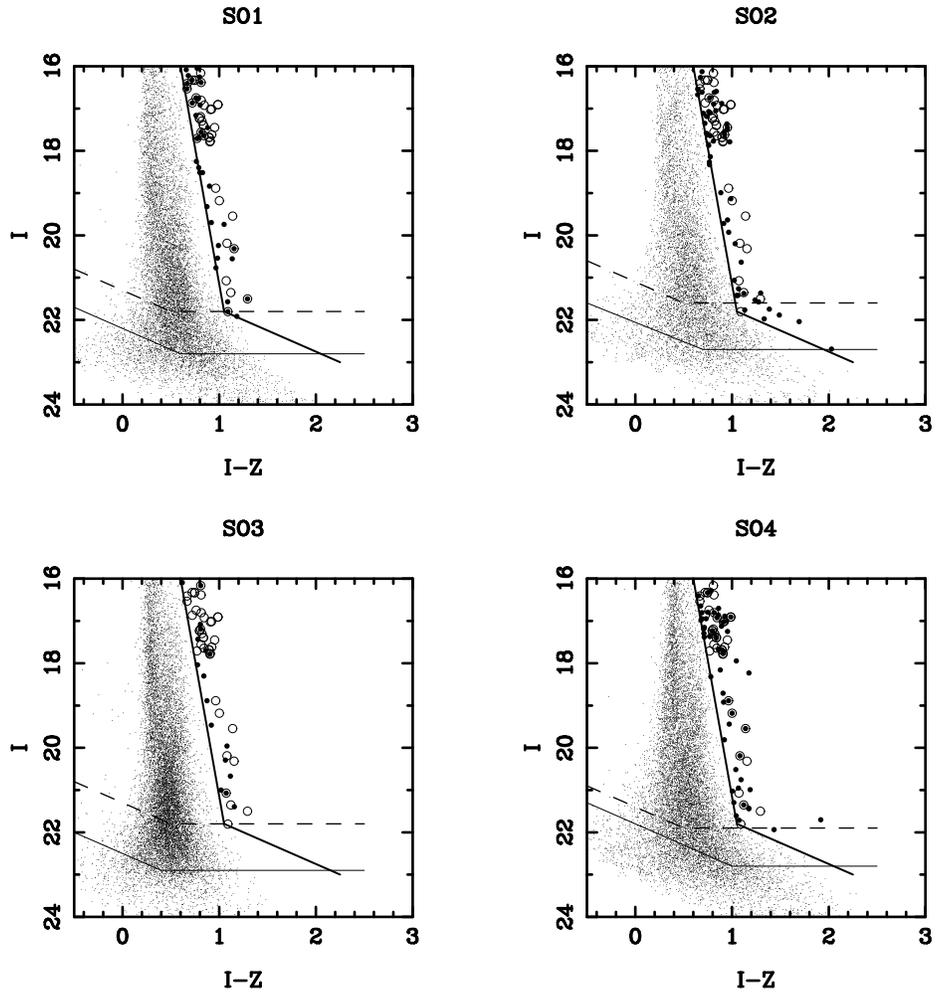}
\caption[]{\label{fig2}$I$ vs. $I-Z$ color--magnitude diagrams for the four pointings. 
The selected objects are indicated by solid circles, while those previously 
confirmed members are represented by open circles.  The completeness and limiting 
magnitudes are also shown in dashed and solid lines, respectively. The lower envelope for candidates selection is indicated by a thicker solid line.}
\end{figure}

We have constructed an $I$, $I$-$Z$ color--magnitude diagram for each field in order to select 
the true cool cluster member candidates. Figure \ref{fig2} shows 
the sum of all $I$, $I$-$Z$ diagrams for each field. We have selected 158 objects 
from these color--magnitude diagrams. They have magnitudes in 
the range $I$=16--23\,mag, which corresponds to a mass interval from 0.1 down to 0.008 
M$_{\odot}$ at the most probable age of $\sigma$ Orionis (2--4 Myr). Five of them are repeated since they have been selected 
twice in different detectors. 
These 153 objects show brighter magnitudes and redder
colors than field objects and follow the photometric sequence of cluster members found in 
previous studies. The lower envelope for candidate selection that separates field objects and cluster members is indicated in Figure \ref{fig2}. Of the 153 selected 
objects, 77 had been previously identified by other surveys, while 76 are reported here for the first time. 
The distributions of these objects can also be 
seen in Figure \ref{fig1}, where these are indicated by stars. 

The technique of low mass member selection from color--magnitude
diagrams based on optical filters, such as $R$, $I$ and $Z$, has been successfully used in young nearby
clusters to distinguish them from background objects (\citealt{proser94,osorio99b}; BZOR ; \citealt{bouvier98}).
The most important sources of contamination in these surveys are field M dwarfs. Bright galaxies are expected to be 
mostly resolved and given the galactic latitude of the $\sigma$ 
Orionis cluster (b=--17.3 deg), giant stars are not expected to contribute in  significant numbers
 ($<$ 5\,\%) in comparison with
main-sequence dwarf stars \citep{kirketal94}. 
 According to the density of M field dwarfs obtained by \cite{kirketal94} and \cite{cru03}, the density of early and 
mid-L field dwarfs obtained by \cite{kirk00} and absolute magnitudes derived by \cite{vrba04}, 
we expect that our photometric sequence for the cluster is contaminated by about 16 late M spectral 
type dwarfs and 1 early to mid-L field dwarf within the completeness of our survey (see more details in Table \ref{tab5}). 
This result is consistent with similar estimations made by \cite{cabetal08a}. 
Background stars reddened 
by interstellar extinction and unresolved galaxies could also populate the optical photometric sequence of the cluster. In this case 
additional selection criteria are necessary to distinguish bona fide cluster members from these
contaminants. 
The combination of optical and infrared data  has proved to be a fiducial technique 
to distinguish bona fide cool cluster members from background objects (\citealt{osorio97a,osorio97b,martin00}; BMZO). 
The membership of most of the low mass stars and brown dwarfs ($>$ 90\,\%) identified using both optical and infrared 
photometric sequences, in low-extinction clusters like the Pleiades and $\sigma$ Orionis, was later confirmed 
from proper motions, radial velocity or the presence of lithium (\citealt{osorio97b}; \citealt{moraux01}; 
\citealt{kenyon05}; \citealt{bihain06}; \citealt{cab07a}).

\begin{figure}[]
\epsscale{0.6}
\vspace{1.75in}
\plotone{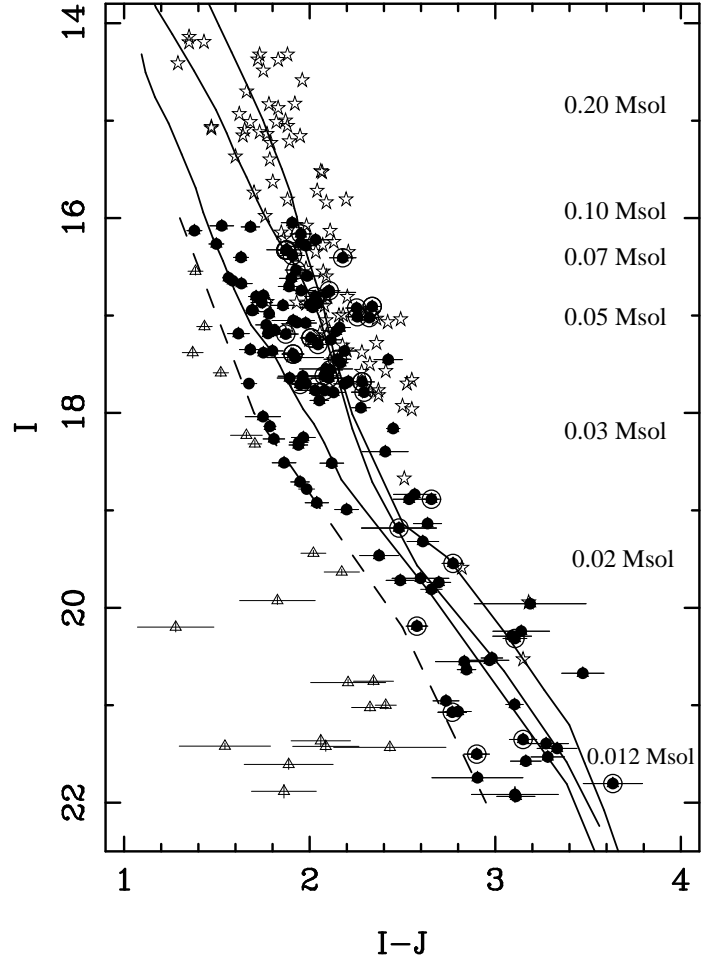}
\caption[]{\label{fig3}$I$ vs. $I-J$ color--magnitude diagram. Open stars represent objects from previous surveys. 
Solid circles indicate good cluster member candidates following the infrared photometric sequence of the cluster, while those previously 
confirmed members are also represented by open circles. Open triangles denote likely non-cluster members. Dashed line indicates the lower envelop of bona fide members.
The solid lines represent the 1, 3 and 10 Myr isochrones from the Lyon group. Masses for the 
3 Myr iscochrone are also indicated.
}
\end{figure}

Figure \ref{fig3}  represents a $I$, $I$-$J$ diagram with cluster member candidates from previous surveys (BZOR, 
\citealt{bejar04a}), indicated by open stars, and the 143 objects  with available $J$-band photometry from the present 
survey, represented by solid circles and open triangles. All of them are within the completeness magnitude of the survey 
($I$=16--22\,mag). According to evolutionary theoretical models, this corresponds to a mass interval from 0.1 down to 0.013 
M$_{\odot}$.    
It can be seen from Figure \ref{fig3} that 124 candidates (solid circles) of the 151 selected objects in the 
optical diagrams within the completeness magnitude show redder colors and magnitudes brighter than the 
lower envelope of the photometric sequence of previously confirmed members, which 
roughly corresponds to the 10 Myr isochrone; we will thus consider these objects as the 
likely photometric cluster member candidates of the present survey. 
There are 27 objects, within the completeness of the survey, that 
 present bluer $I-J$ colors than expected for the photometric sequence of the cluster in 
Figure \ref{fig3}; these will be considered as probable non-members in the rest of the paper. 
The full list of objects presented in this paper is given in Table \ref{tab2}. Their membership status is also indicated in the last column of Table \ref{tab2}. 
There are also two objects with $I\ge$\,22\,mag for which there are no available infrared data deep enough to restrict their 
belonging to the infrared photometric sequence. 
Since they are not within the
completeness magnitude of our survey, we will not consider them for the analysis of the 
very low mass stars and brown dwarfs of 
the cluster in next sections.

\section{Spatial distribution of the very low mass stellar and substellar population}

In this section we analyze the spatial distributions of the very low mass stars and brown 
dwarf candidates selected in the present survey. We have selected 124 good cluster member candidates 
within the completeness magnitude that follow both the optical and infrared photometric sequence of 
previously known low mass members of $\sigma$ Orionis. 

\subsection{The center of the $\sigma$ Orionis cluster}
\begin{figure}[]
\epsscale{0.7}
\vspace{1.75in}
\plotone{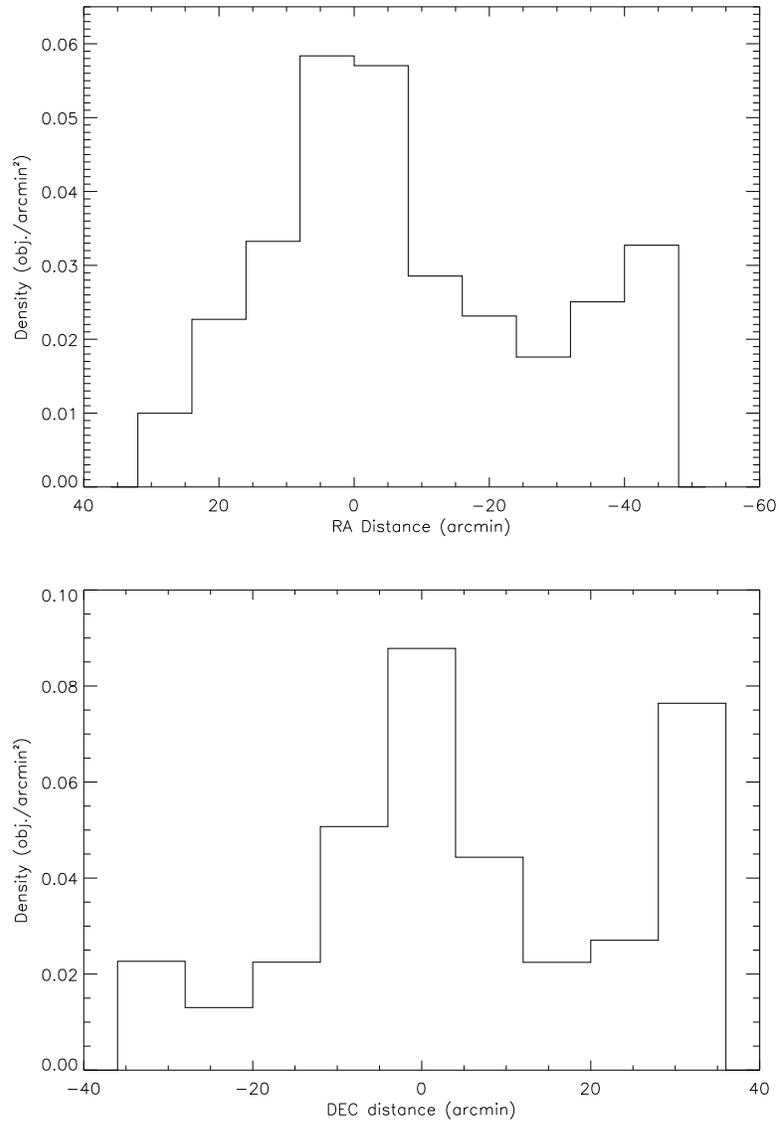}
\caption[]{\label{fig4}Projected surface density  of the substellar population vs. the separation 
in $\alpha$ and $\delta$ to 
the center of the cluster. The presence of possible contaminants from the $\zeta$ Orionis cluster 
and the Orion background population can be seen to the north and west of the $\sigma$ Orionis cluster.}
\end{figure}

The first question arisen in our study is whether there is clear evidence of the existence 
of a clustering of substellar objects around the multiple star $\sigma$ Orionis. The representation of the spatial 
distribution in Figure \ref{fig1} shows that there is a concentration of substellar objects around 
$\sigma$ Orionis. To test this we have calculated the distributions of the object density per arcmin$^2$ 
in the present survey along the $\alpha$ and $\delta$ axis 
centered on the $\sigma$ Orionis AB coordinates. 
A representation of these distributions can be seen in the top and bottom panels of Figure \ref{fig4}. 
It can be seen that 
the distribution decreases from the central star in both the $\alpha$ and the $\delta$ axis, indicating 
the existence of a greater concentration of objects around $\sigma$
Orionis. It is interesting to note that from both figures we can see that there 
is also an increase in the last bins (at separations larger than 30 arcmin) to the north and west of $\sigma$ Orionis. 
Some of these objects are located at distances closer to the star $\zeta$ Orionis and could 
be related to the existence of a substellar population around this star. 
The rest of these objects are located closer to $\sigma$ Orionis than to any other OB star, 
but this population can be related 
to the sparse, wide clustering around the  $\epsilon$ Orionis cluster \citep{sherry03,briceno05,cabysol08}. 
It might be that the population identified by \cite{jeffries06}, which is kinematically different from the  $\sigma$ Orionis cluster 
and seem to be concentrated to the north-west of the star, is also related with the excess of sources observed.  
Objects in this region are identified in the last column of Table \ref{tab2}. After discarding this population of possible Orion background objects 
a total of 102 bona fide member candidates remains concentrated around the $\sigma$ Orionis star.  
We determine that the central coordinates of the cluster, according 
to the substellar distribution, are within the central bins of 5 arcmin around 
the massive star. This estimate is not very precise because of the limits 
imposed by the geometry of our survey and  the radial dependence of the object 
density in a cluster. 
The center of mass of the multiple star $\sigma$ Orionis is located a few arcseconds from
the central star $\sigma$ Orionis AB (see \citealt{cab07,cab08b}). We can estimate the center of mass
of the substellar population of the cluster and find that this is located within 
2 arcmin of the more massive stars; considering the uncertainties, we can argue that they are the same. 
In the remaining discussion of the spatial distribution 
we consider the coordinates of $\sigma$ Orionis AB as the central coordinates of the cluster. 

\subsection{The radial distribution of very low mass objects around $\sigma$ Orionis}

\begin{figure}[]
\vspace{1.75in}
\plotone{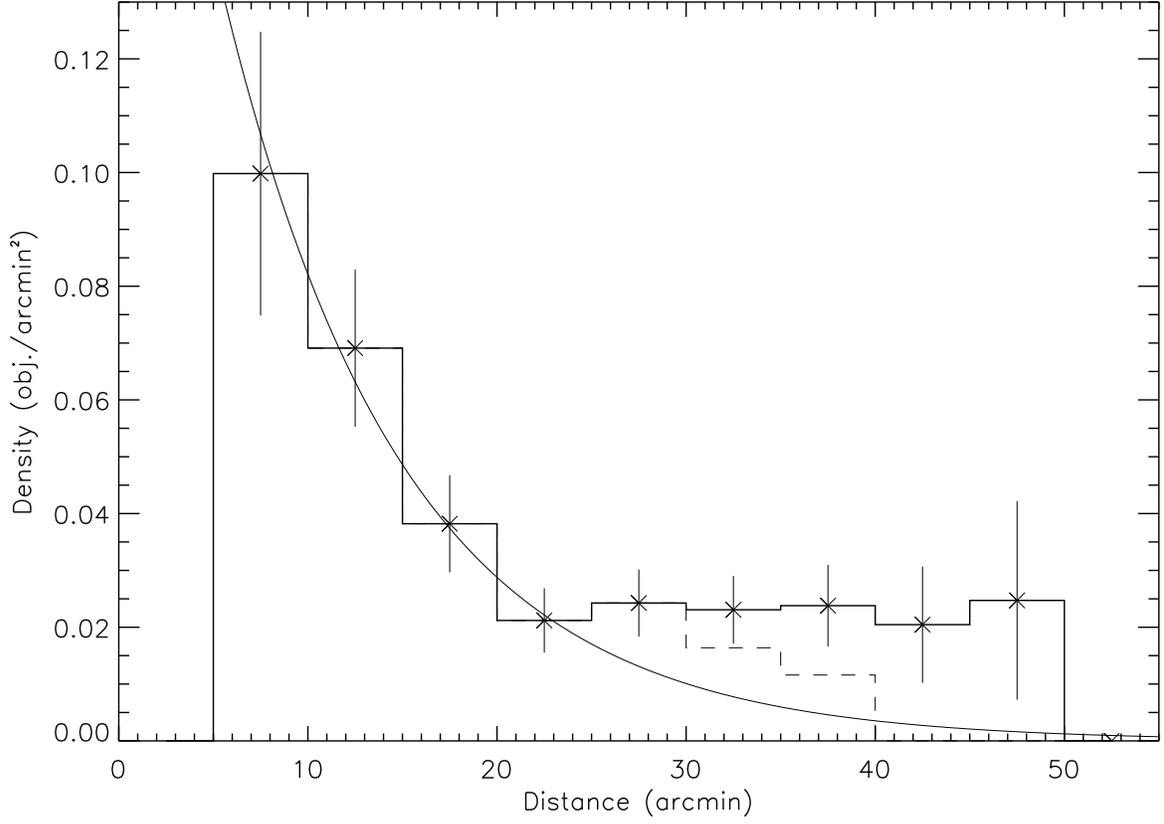}
\caption[]{\label{fig5}Projected spatial distribution of the substellar population. The best exponential law 
fit ($\sigma=\sigma_{0}\,e^{-r/r_{0}}$, where $\sigma_{0}$=0.23\,obj/arcmin$^2$ and $r_{0}$=9.5\,arcmin) 
is also indicated (see text for details). 
The dashed line histogram at distances larger than 30 arcmin marks the spatial distribution of 
likely $\sigma$ Orionis cluster candidate members only (after substracting possible Orion background population contaminants).}
\end{figure}

The next objective in our study is to characterize the radial distribution of the surface 
density of the substellar objects in the cluster. In order to calculate this we have estimated 
the surface density of the number of objects in concentric coronas at different distances to 
the center of the cluster. This is a decreasing function with distance to the center and 
is shown in Figure \ref{fig5}. The increase in this distribution at distances larger than 30 arcmin can be explained 
by the presence of other population of Orion (see previous subsection). 
We can try to adjust this distribution with different empirical 
functions obtained for star clusters such as an exponential \citep{van60} or a King distribution \citep{king62}. 
In the case of the $\sigma$ Orionis cluster, due to the sharpness of the decay and the contamination of a 
background population at large distances, 
the former seems to be more suitable. In addition, the King function was developed to reproduce dynamically 
relaxed clusters, and, 
given the very young age of $\sigma$ Orionis, we do not expect that dynamical evolution is yet important.   
We have adjusted the surface density to an exponential law,   
$\sigma = \sigma_{0} e^{-r/r_{0}}$,
where $r_{0}$ is the characteristic radius of a cluster, defined as the radius 
at which the density has diminished by a factor $e$, and $\sigma_{0}$ is the central
density. The best fit values obtained for these variables in the distance range between 5 and 25 arcmin from the center are: 
 $r_{0}$=9.5 $\pm$ 0.7 arcmin  ($\sim$ 1 pc at the distance of $\sigma$ Orionis), and  $\sigma_{0}$=0.23 $\pm$ 0.03 object/arcmin$^2$. Similar results are found for brighter candidates ($I<$18\,mag) 
and fainter ones ($I>$18\,mag), indicating that there is no significant mass segregation between more and less massive brown dwarfs in the cluster. 
The dashed line in Figure \ref{fig5} shows the surface density of the 102 $\sigma$ Orionis member candidates after 
subtracting the possible background population located at projected distances larger than 30 arcmin to the north and west of the cluster. 
We can see that the estimated exponential law extrapolates 
reasonably 
well this surface density to distances up to 40 arcmin, although we can see that the real numbers could be 
slightly larger. 
This could indicate a lower decrease in the surface density at these distances as pointed out
by \cite{cab08b}, who argued that 
an extended halo of low mass stars is present in the cluster at distances larger than 20 arcmin. 

Although comparison with previous studies of spatial distribution of the low mass population of the cluster 
is difficult due to the different functions adopted to fit the surface density, we can say that the radial distributions 
of substellar objects found in this paper is similar to that of the low mass stars of the cluster \citep{sherry04,cab08b}. 
These two studies used a King model to adjust radial surface density and found a core radius $r_{c}$ of 
10--12 arcmin ($r_{0}\sim$ 1.36 $r_{c}$). In addition, \cite{cab08b} also studied an exponential and a potential law and found a characteristic 
radius between 12 and 18 arcmin 
(see their figure 3) and find that the best fit for the radial surface density is obtained for an $r^{-1}$ function. 
Another study that has investigated the spatial distribution of low mass stars and brown dwarfs in 
the $\sigma$ Orionis cluster is that by \cite{lodieu09}. In this case they did not adjust any function to the radial
distribution, but they found that cluster members are clearly concentrated within the central 30 arcmin;
 they also found a deficit 
of brown dwarfs in the central 5 arcmin with respect to the stars. A similar result was also previously 
reported for very low mass stars and 
brown dwarfs by \cite{cab08b}. 
Our present survey does not cover this region and hence we cannot 
address this question. Preliminary results of the spatial distribution of brown dwarfs using the same optical
 data as those studied here were also
presented in \cite{bejar04b}. They found similar results of the central density and characteristic radius 
within 1--2-$\sigma$ the quoted uncertainties. 

\subsection{Sub-clustering or aggregations of very low mass objects in the $\sigma$ Orionis cluster}

To investigate whether the location of brown dwarfs in the $\sigma$ Orionis cluster can be 
affected by the presence of other substellar objects, which could be a residual phenomenon 
of a common process of formation, we have investigated whether the distribution of the substellar 
population is Poissonian at each radial distance to the center of the cluster. 
In order to do that, we have investigated the distance to the nearest neighbor. 
We have compared the results obtained in our sample of substellar objects 
with theoretical predictions for a Poissonian distribution and with the results for 
Monte Carlo simulations of clusters in the same conditions as ours. 
We computed 10000 simulated clusters with the same radial surface density of 
substellar objects obtained in Section 4.2 for separations to the center  
between 0 and 50 arcmin and with a total number of 130 objects, given by the integration 
of this exponential distribution between both radii. An increase in the number of 
simulated clusters does not improve the statistical uncertainties in our computations
because these are dominated by the small number of objects (between 10 and 20) in the radial 
bins of the distribution.

\begin{figure}[]
\vspace{1.75in}
\plotone{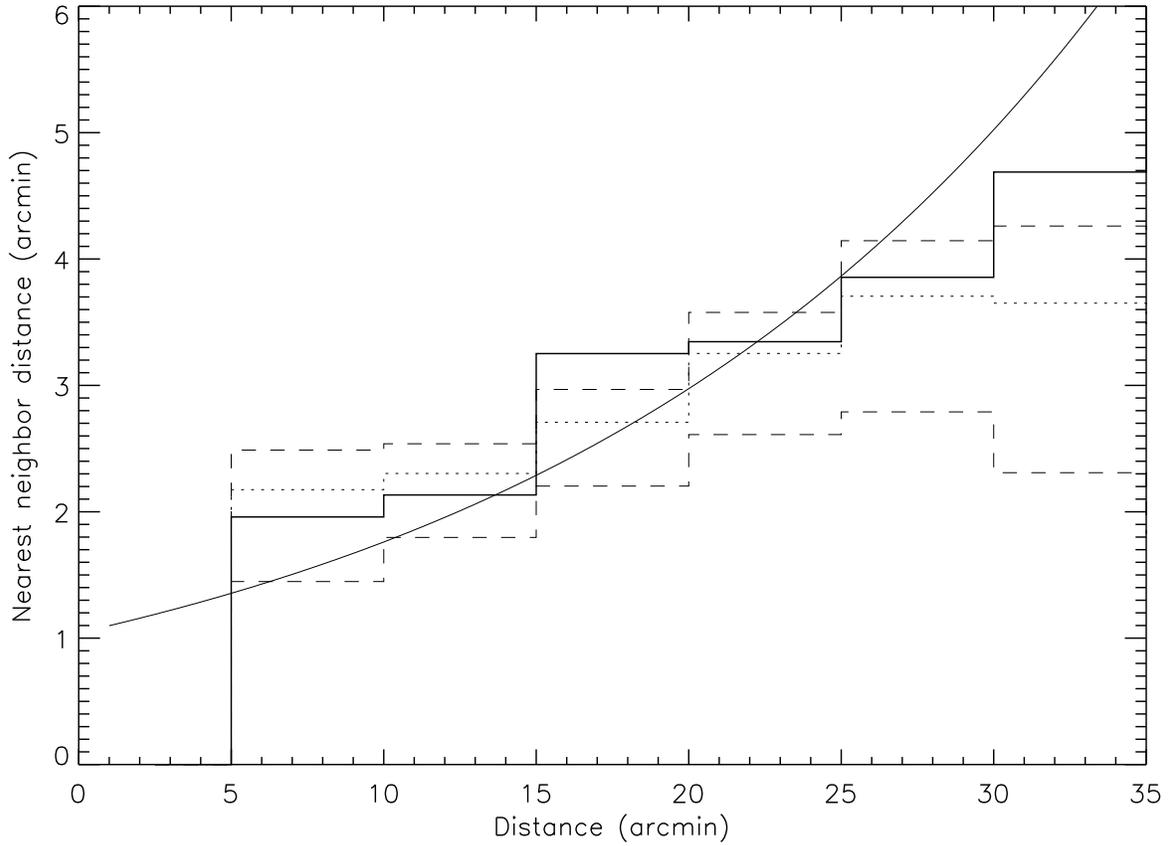}
\caption[]{\label{fig6} Average nearest neighbor distance between $\sigma$ Orionis candidates (solid histogram) as a function of distance to the center 
of the cluster. For comparison purposes, we also show the mean values (dotted line) and the first and third quartiles (dashed line) of the same estimator 
obtained from a Monte Carlo simulation of 10,000 clusters with the same radial distribution and area coverage as our survey, and the expected mean 
values from a Poissonian distribution (solid line) with the same radial surface density as the cluster.}
\end{figure}

\begin{figure}[]
\vspace{1.75in}
\plotone{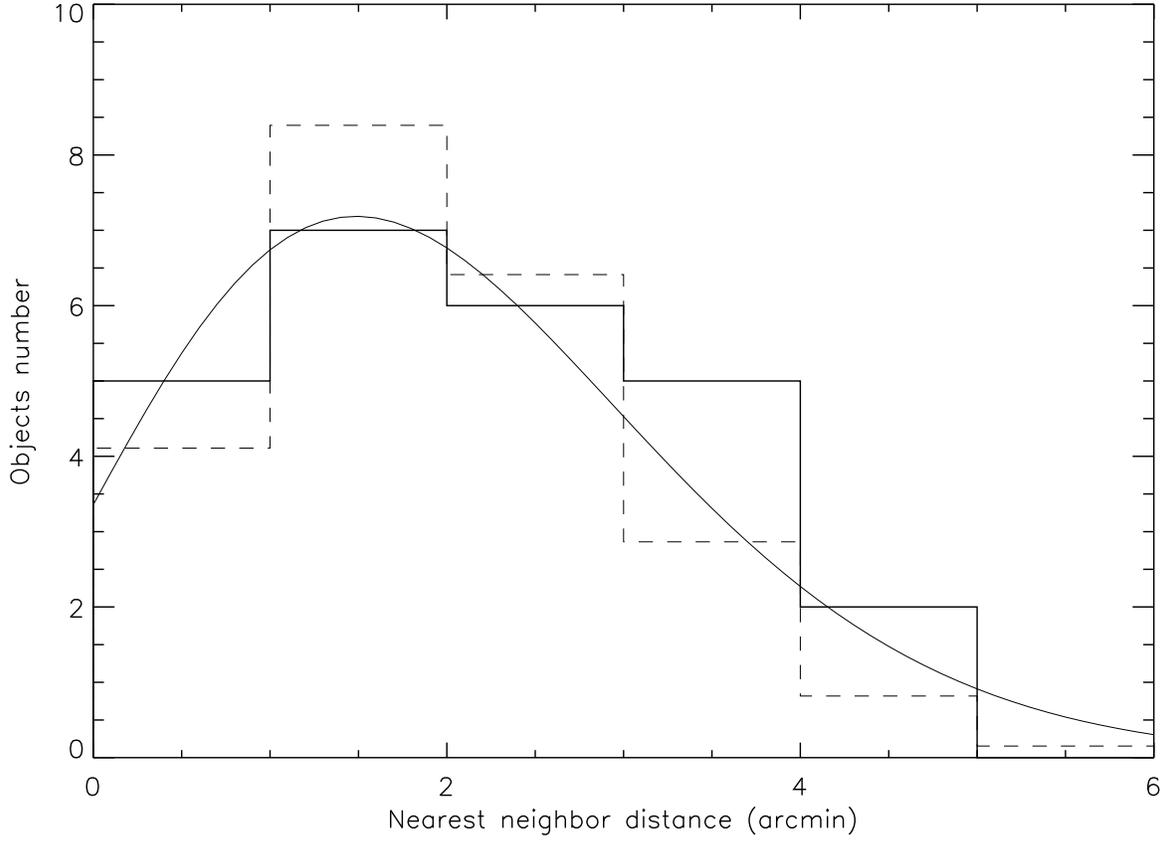}
\caption[]{\label{fig7} Nearest neighbor distances of good cluster member 
candidates (solid histograms) at separations between 10 and 15
arcmin from the center. For comparison purposes, we also show the  histogram  of the values of simulated clusters 
(dashed line) and the expected Poissonian distribution (solid line) with the mean surface density of the cluster at these distances.}
\end{figure}

The solid histogram of Figure \ref{fig6} shows the average value of the nearest neighbor distance of the $\sigma$ Orionis member candidates  
as a function of the separation from the cluster centre. For comparison purposes, we also show 
the expected value for a Poissonian distribution and the nearest neighbor distances derived from Monte Carlo simulations. In this case we have only taken into 
account those objects in the simulated clusters that are located within the surveyed area of the real observations. The reason 
is to eliminate border effects due to the particular geometry of our study that might affect the nearest neighbor distance 
estimator. 
From this Figure we can see that this distance is similar in our sample to that estimated 
for the Monte Carlo simulations, and that at close separations to the center it is similar to that expected from a 
theoretical 
Poissonian distribution. Between 5 and 15 arcmin both the nearest neighbor distance of the sample and simulations are 
slightly larger than the Poissonian homogeneous distribution due to boundary problems caused by the incompleteness 
of our survey at these distances. This effect is not a consequence of the existence of sub-clustering as shown by the
simulations because, if this happened, we would expect lower separations. At separations larger than 30 arcmin 
we are dominated by the 
incompleteness of our survey and separations are lower than 
expected for a Poissonian homogeneous distribution. As an example, the solid histogram of Figure \ref{fig7} show the nearest 
neighbor distances of the candidates located between 10 and 15 arcmin from the cluster centre. For comparison, we also show the  histogram  of the 
values of simulated clusters and the expected Poissonian distribution with the mean surface density of the cluster at these distances. 
We can say that all these distributions are consistent among themselves, considering the uncertainties due to the low number of objects per bin. 
These results indicate that the distribution substellar objects at the same radial distance to the center of the cluster 
is almost Poissonian and, hence, that
this is not dominated by the presence of aggregations or sub-clustering of these objects.

\section{Infrared excesses}

\subsection{Near-infrared excesses}

\begin{figure}[]
\epsscale{0.6}
\vspace{1.75in}
\plotone{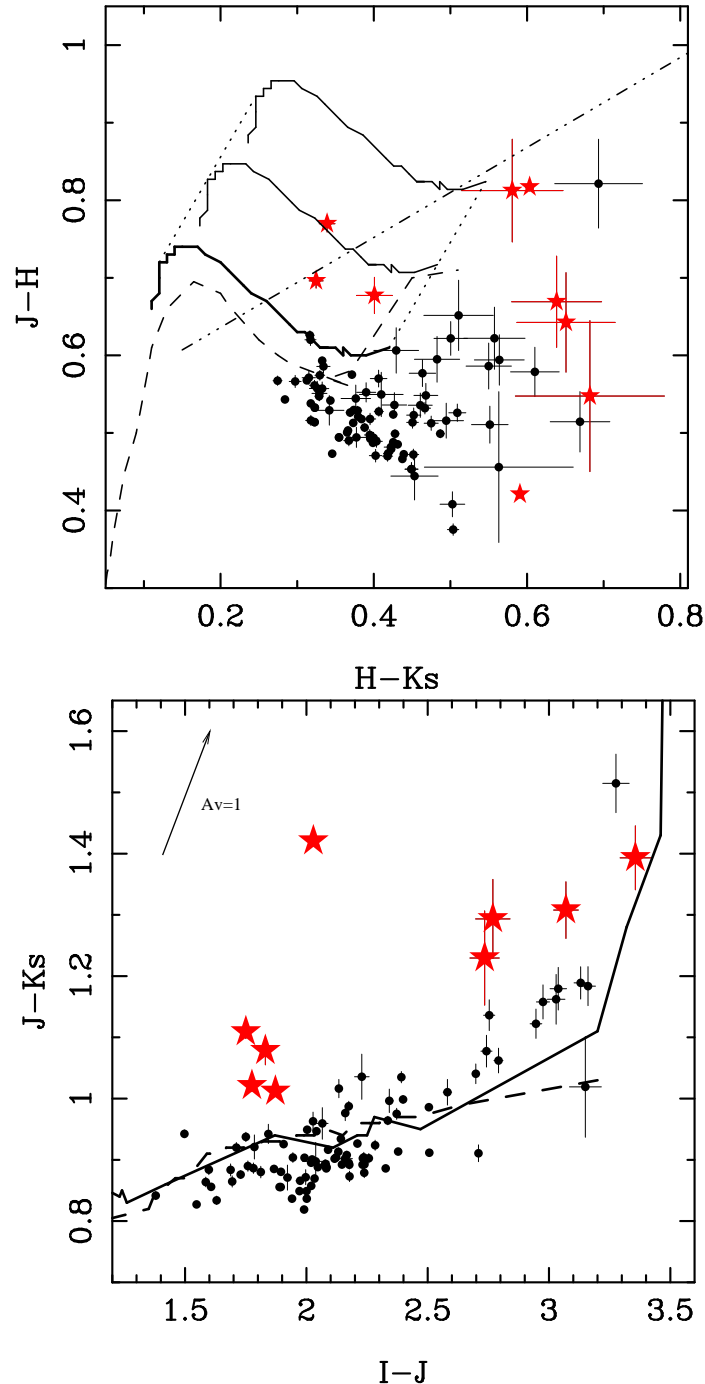}
\caption[]{\label{fig9} Upper panel: $J-H$, $H-K_{s}$ color--color diagram of 
selected candidates (solid circles) with available 
$JHK_{\rm s}$ photometry from UKIDSS. Objects with near-infrared 
excesses are indicated by stars and red colors (in the electronic version). 
The 3 Myr Next Gen isochrone from the Lyon group \citep{baraffe98} 
reddened with visual extinctions of $A_{V}$ = 0, 1 and 2 are plotted in solid lines from bottom to top (joined by dotted lines). 
The field dwarf sequence (dashed line) from \cite{bessell88} and \cite{kirk94} 
and the classical T Tauri star loci (dash-dotted line) from \cite{meyer97} are also indicated. 
Lower panel: $J-K_{s}$, $I-J$ color--color diagram of selected candidates (solid circles) with available 
$JHK_{\rm s}$ photometry from UKIDSS. Objects with near-infrared 
excesses are indicated by stars and red colors (in the electronic version). 
The field M dwarf sequence (solid line) from \cite{bessell88} and \cite{kirk94} 
and the field M--L dwarf sequence from \cite{leggett02} (dashed line) are also indicated. 
}
\end{figure}

 We have used the available $JHK_{s}$ photometry from the UKIDSS catalogue of those reliable cluster member candidates with 
errors smaller than 0.10\,mag to study the presence of near-infrared excesses. 
A total of 98 objects have available photometry within this requirement, which includes the majority of our candidates 
within the completeness of the survey. 
According to theoretical evolutionary models, these cluster member candidates  
have masses in the interval 0.11--0.013 M$_{\odot}$.
 We have built the $H-K_{s}$, $J-K_{s}$ and  $I-J$, $J-K_{s}$ color--color diagrams shown in Figure \ref{fig9}. 
From these representations we can see that 5 objects present near-infrared $H-K_{s}$ and $J-K_{s}$ colors 
deviating from the expected sequence of non reddened cluster members by more than 2-$\sigma$. 
All except one and another 4 of the candidates show a redder $J-K_{s}$ color than expected for 
their $I-J$ color by more than 2-$\sigma$.   
These objects are marked in the last column of Table \ref{tab3}.  
This indicates that 5--9 of the 98 $\sigma$ Orionis cluster member candidates with accurate UKIDSS photometry ( $\sim$ 5--9\,\%) could have near-infrared excesses. 
Three of these objects (\object{S\,Ori\,J053849.29--022357.6}, \object{S\,Ori\,J053907.56--021214.6}, and  \object{S\,Ori\,J053940.58--023912.4}) also show near-infrared excesses   
using the photometry from 2MASS, while the rest of them are very faint or have large error bars in this catalogue. 

 Using the same criteria as above, we have found that two or three out of the 17 objects (12--18\,\%) 
of the Orion background population have also these excesses, which indicates a similar fraction to that of the 
cluster member candidates.  Curiously, 2--3 out of the 9 objects (22--33\,\%) considered as non-members present near-infrared excesses. 
As explained before, we estimated that most of the non-member contaminants of our survey are formed by 
field M dwarfs; hence, it is not expected that they still have a disk and have such a large fraction of infrared excesses.  
These three objects (\object{S\,Ori\,J053945.35--025409.0}, \object{S\,Ori\,J053840.38--030403.2}, and \object{S\,Ori\,J053936.08--023627.3}) 
lie quite close to the photometric sequence of the cluster and hence might be members. In these cases, likely variability and photometric errors can explain 
their location in the $I$, $I$-$J$ color-magnitude diagram. Because they show infrared excesses, they could also be members partially occulted by the presence of their disks. 
Another possibility is that these three objects could be unresolved galaxies; some of
these galaxies show very red near and mid-infrared colors but not so red $I-J$ colors (see, for example, \citealt{cabetal08b}).  
Additional photometry and spectroscopy is required to determine the true nature of these objects. 
If they are really bona fide members of the cluster, the fraction of substellar objects with near-infrared excesses 
could be increased by up to 7--12\,\%. We will not considere these objects as cluster members for our estimation of the substellar mass function 
in next section. Following these considerations, we have estimated that about 2--3\% of the cluster members could be lost with our selection criteria. 
 
In summary, we have found that $\sim$ 5--9\,\% of the very low mass stars and brown dwarf candidates in 
the $\sigma$ Orionis cluster have near-infrared 
excesses at 2.2\,$\mu$\,m that could be related to the presence of disks. Other studies have found similar results with a 
fraction of near-infrared excesses 
between 5 and 12\,\%  in cluster member candidates \citep{oliveira02,barrado03,bejar04a,cab08a}.
 
\subsection{Mid-infrared excesses}

\begin{figure}[]
\vspace{1.75in}
\plotone{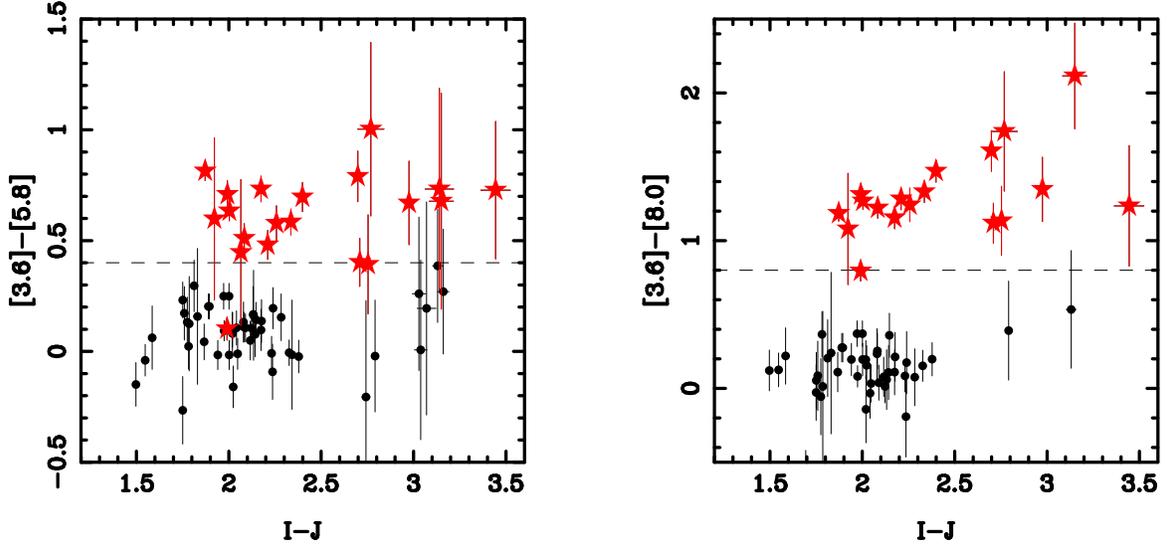}
\caption[]{\label{fig10} Left panel: $[3.6]-[5.8]$, $I-J$ color--color 
diagram of candidates (solid circles) with available 
photometry from IRAC/{\em Spitzer}. Objects with mid-infrared excesses are 
indicated by stars and red colors (in the electronic version). 
Right panel: $[3.6]-[8.0]$, $I-J$ color--color diagram of candidates with available 
photometry from IRAC/{\em Spitzer}. Symbols are the same as in the left panel. 
The horizontal dashed lines indicate our color criteria to identify flux excesses.}
\end{figure}

In order to identify the population of substellar objects with mid-infrared excesses in the $\sigma$ 
Orionis cluster, we have 
used the available 3.6, 4.5, 5.8 and 8.0\,$\mu$\,m photometry from the IRAC/{\em Spitzer} data. 
Following a similar procedure as in \cite{cab07a}, we have built $[3.6]-[5.8]$, $I-J$ and  $[3.6]-[8.0]$, $I-J$ 
color--color diagrams, shown in Figure \ref{fig10}. A total of 67 good cluster member candidates have $I$, $J$, $[3.6]$, $[5.8]$ or 
$I$, $J$, $[3.6]$, $[8.0]$ photometry. 
Nineteen of the 67 objects (28$\pm$7\,\%) show $[5.8]-[3.6] \geq$ 0.4\,mag and 
18 of the 58 (31$\pm$7\,\%) show $[3.6]-[8.0] \geq$ 0.8\,mag. We have used these color criteria to identify mid-infrared flux excesses, which are 
based on the separation between objects with and without disks adopted in \cite{cab07a}, \cite{osorio07}, and references therein. 
The presence of mid-infrared excesses is also indicated in the last column of Table \ref{tab3}.
All the candidates that show infrared excess in the $[5.8]$ band and have available photometry in $[8.0]$ also present  
infrared excess in this band. In addition, all except one of the objects  
that show infrared excess in $[8.0]$ also show it in $[5.8]$. 
By considering both the excesses at $[5.8]$ and $[8.0]$, we have found that 20 of the 67 candidates (30$\pm$7\,\%) show a larger emission at wavelengths 
longer than 5.8\,$\mu$\,m.
In any case all these results are lower limits, since some of the candidates can be field dwarf contaminants. 
Adopting the maximum number of contaminants previously estimated, this fraction could be increased up to 40$\pm$9\,\%.    
Six of the 9 cluster member candidates with near-infrared excesses have available IRAC/{\em Spitzer} photometry in 
$[3.6]$ and $[5.8]$ or $[8.0]$. Two of them also have a larger emission in the mid-infrared, while the other four have not. 
All except two (18) of the candidates presenting mid-infrared excesses have available accurate UKIDSS photometry. 
Only the two objects mentioned above also show a larger emission in the $K$-band, while the rest of them (16) only present excesses at wavelength longer than 5.8\,$\mu$\,m.

In summary, we have found that about 30$\pm$7\,\% of the substellar cluster member candidates 
show mid-infrared excesses that are probably associated with  the 
presence of disks. A similar result has also been found in other studies of the cluster, both in low mass stars, 
brown dwarfs and
planetary-mass domain  \citep{jaya03,oliveira04,oliveira06,hernandez07,cab07a,osorio07,scholz08,luhman08}. Given that the fraction 
of candidates with excesses at wavelengths longer than 5.8\,$\mu$\,m are notably larger than at 2.2\,$\mu$\,m,   
most of the disks around these objects seem to be ``transition disks'',  which are characterized for the weak or absent 
near-infrared emission and the presence of mid- and far-infrared excesses \citep{strom89,sicilia06}. This is explained by a 
process of clearing in the inner part of the disk, probably due to the formation of planetesimals, leading to a transition from optically 
thick to optically thin inner disks.

\section{The substellar mass function}

\index{mass function}

In this section we estimate the mass function in the surveyed area of the 
$\sigma$ Orionis cluster from low mass stars until the deuterium burning  
mass limit. Instead of investigating the mass function we are going to study the mass 
spectrum, defined as the differential frequency distribution of stellar masses, $\phi(m) = dN/dm$, 
(in this form $\phi(m)$ $\sim$ m$^{-\alpha}$, and the Salpeter exponent is $\alpha$=--2.35; \citealt{salpeter55}). 
In order to derive the mass spectrum or the mass function in clusters, it is necessary to obtain 
a luminosity function from the observations and a mass--luminosity relationship, which can be provided 
by observations or by theoretical models. The advantage of having a number of 
objects with similar age is that there is no need to make any assumption about the star 
formation history, a question that affects the solar neighborhood and field studies. In addition to this, 
for stellar clusters a unique mass--luminosity relationship is needed, which is important in 
the substellar range, where this
function depends drastically on time.

\begin{figure}[]
\vspace{1.75in}
\plotone{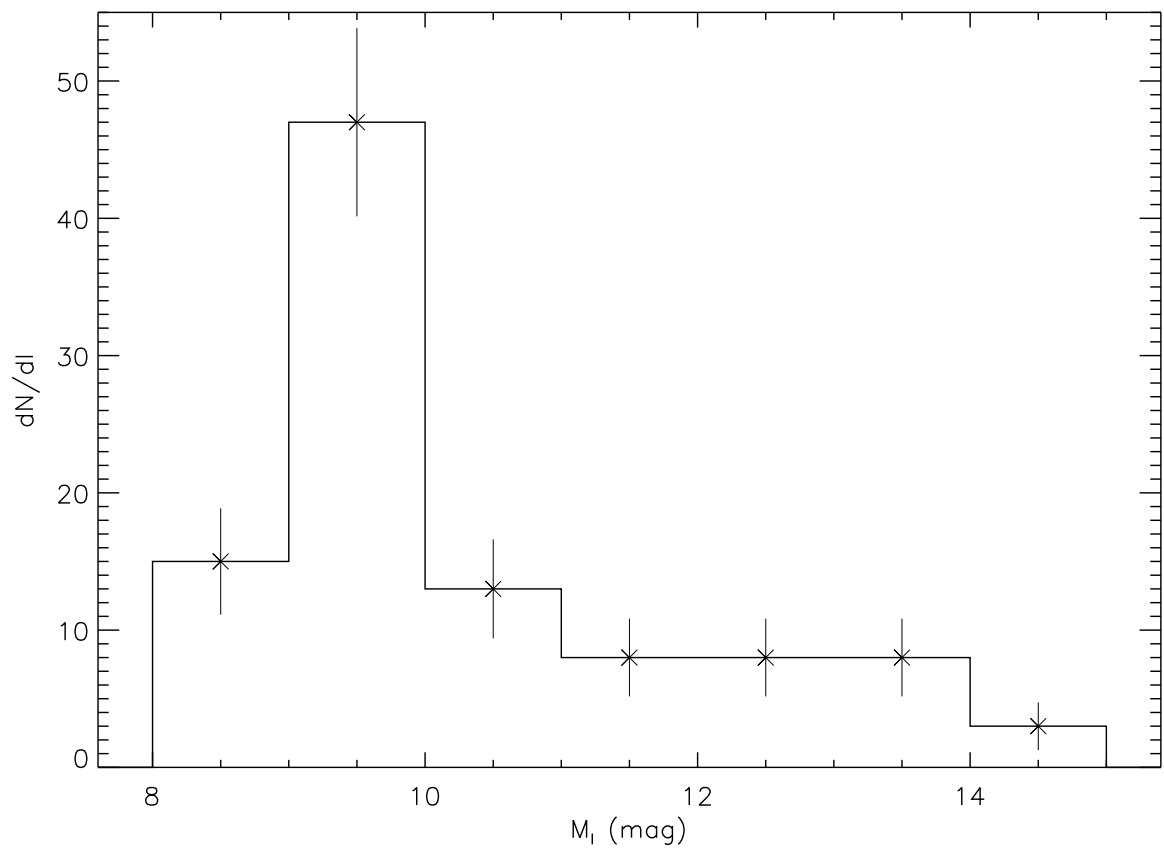}
\caption[]{\label{fig11}Luminosity function of the selected candidates. A histogram of the number of objects 
in each bin of magnitude is represented.  }
\end{figure}

\begin{figure}[]
\epsscale{0.6}
\vspace{1.75in}
\plotone{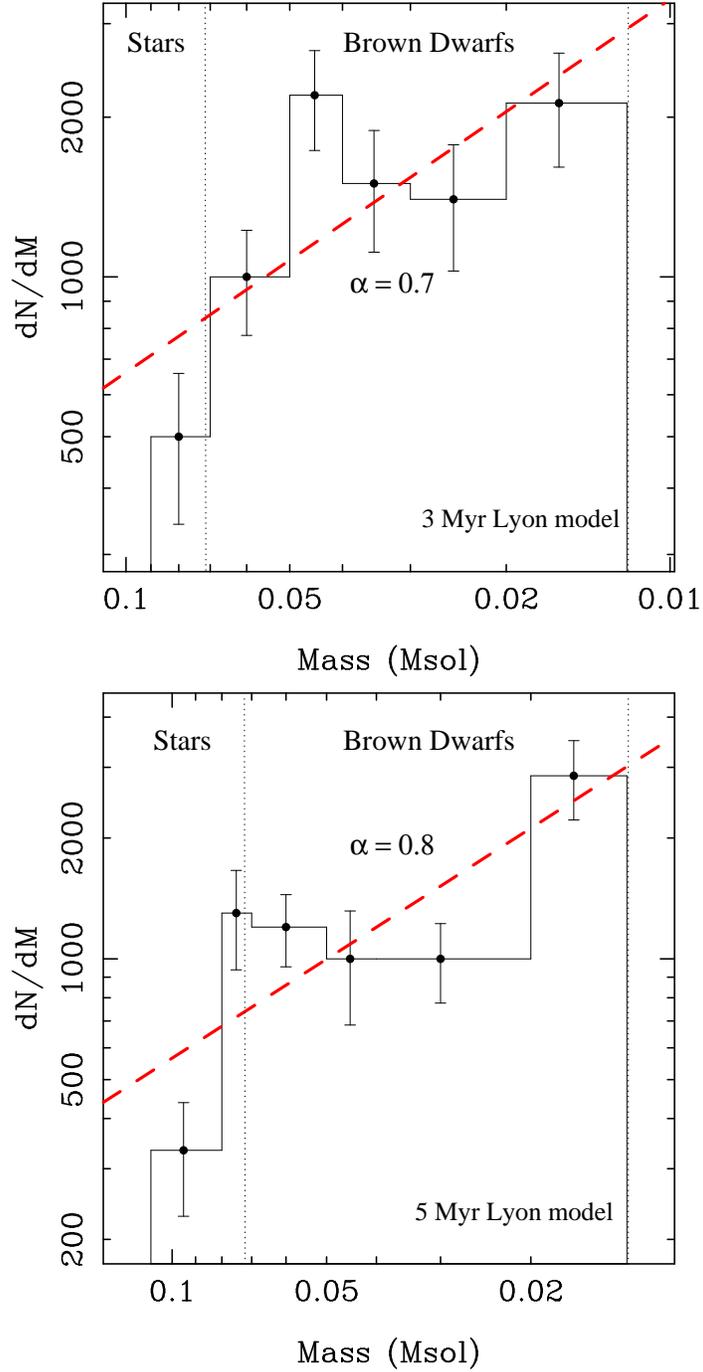}
\caption[]{\label{fig12}The mass spectrum of $\sigma$ Orionis. Only bona fide cluster member
 candidates within the completeness 
magnitude of present survey have been considered. 
The theoretical isochrone of 3\,Myr (top panel) and 5\,Myr (bottom panel) from the Lyon Group (1998) has been used to derive masses. 
Histograms represent the number of objects per interval of mass ($dN/dm$). The best fit to a potential 
law ($dN/dm$ $\sim$ m$^{-\alpha}$)
with an exponent of  $\alpha$=0.7 for 3\,Myr and 0.8 for 5\,Myr is represented by a dashed line (in red in the electronic version). }
\end{figure}

The first step in calculating the mass spectrum of the cluster is to define an accurate list of members 
from our observations covering a representative area. Our studied area, although not 
complete over the full extension of cluster, covers the greatest part of it and includes more than 75\,\% of the expected number of cluster members 
in the studied interval of magnitudes ($I$=16--22\,mag), considering the observed surface density distribution.  
The resulting luminosity function obtained from the sum of the 102 bona fide cluster member 
candidates selected from optical and 
infrared photometry is presented in Figure \ref{fig11}.    
To estimate the mass--luminosity relationship, we have 
used theoretical isochrones from several authors: the Arizona group, \cite{burrows97}, 
and the Lyon group, \cite{baraffe98,chabrier00}. We have   
transformed temperatures and luminosities of these models into magnitudes, using 
bolometric corrections, and temperatures and color--spectral type relations from 
\cite{leggett00,leggett02}, \cite{gol04}, and \cite{knapp04}. 
The resulting mass spectra for the Lyon group models at the ages 
 of 3 and 5 Myr are shown in Figure \ref{fig12}. 
In addition, the best fit to a power law ($\phi(m)$ $\sim$ m$^{-\alpha}$) is also shown in the dashed-dotted curve.  
From this figure, we conclude that the mass spectrum is rising in the range from $\sim$ 0.10 M$_{\odot}$ to the 
completeness mass of 0.012--0.013 M$_{\odot}$ with $\alpha$ exponents of  0.7$\pm$ 0.3  
for an age of 3 Myr and 0.8 $\pm$ 0.3 for 5 Myr. Given the uncertainties in the determination of the local density of 
late M and L field dwarfs, we have not carried out any contaminant correction. This would decrease the value of
$\alpha$ index down to 0.5 $\pm$ 0.4, yet consistent with previous measurements within the error bars. 
We have also calculated
 the mass spectrum using Lyon group models for a wider range of possible ages of the cluster within 1--10 Myr and find
that the $\alpha$ index undergoes variations of the order of the error bars in this age interval. 

Previous studies of the mass function of the 
cluster have found similar results. 
For example, BMZO determined this distribution from 0.20 M$_{\odot}$ to 0.013 M$_{\odot}$ 
and found a similar exponent of 0.8 $\pm$ 0.4. \cite{gon06} and \cite{cab07a} studied the substellar mass function 
from the star/brown dwarf borderline down to 0.006--0.007 M$_{\odot}$ 
and found an $\alpha$ index of 0.6$^{+0.5}_{-0.1}$ and 0.6$\pm$ 0.2, respectively. \cite{lodieu09} 
investigated this quantity in a wider range of masses from low mass stars of 0.5 M$_{\odot}$ to the brown dwarf 
domain up to 0.01 M$_{\odot}$ and found a value of 0.5$\pm$ 0.2. More recently, \cite{bihain09} 
have investigated the substellar mass function of the $\sigma$ Orionis cluster in a similar area as studied in 
\cite{cab07a}  up to a few Jupiter masses, and their results suggest a possible turnover 
below 0.006 M$_{\odot}$. 

Many other studies have investigated the substellar mass function, 
the majority of them in clusters or associations, with a few examples in the 
field (see \citealt{reid99,chabrier02,allen05}). 
First estimates of the substellar mass function were made by \cite{osorio97b}, \cite{martin98}, and 
\cite{bouvier98} in the Pleiades cluster. They found that the mass spectrum is still rising below the 
substellar mass limit until 0.040 M$_{\odot}$, while the former find a exponent of 1.0 $\pm$ 0.5, the later 
estimate an exponent around 0.6. 
Other authors have studied the mass function in the nearly complete brown dwarfs regime in $\rho$ Ophiuchi 
\citep{luhman00}, 
IC\,348 (\citealt{najita00}; \citealt{luhman03a}), the Trapezium 
\citep{luhman00,lucas00,hille00}, Chamaeleon \citep{lopez04,luhman07}, 
$\sigma$ Orionis (BMZO; \citealt{gon06,cab07a,lodieu09,bihain09}), $\lambda$ Orionis \citep{barrado04}, 
Upper Scorpius \citep{lodieu07a}, $\alpha$ Persei \citep{barrado02b}, 
Blanco\,1 \citep{moraux07} and the Pleiades cluster \citep{moraux03,bihain06,lodieu07b}. 
All these studies have found a still rising mass
spectrum in the brown dwarf domain and have found $\alpha$ exponents between 0.4--1.0. 
All these determinations are consistent with the results of the present paper and suggest that 
the substellar mass function is similar in various star forming regions with ages below 100--200\,Myr. 
An exception to this behavior is seen in some young star-forming regions such as $\eta$ and $\epsilon$ Cha \citep{luhman04}, 
 NGC~6611 \citep{oliveira09},   and Taurus, where a paucity of brown dwarfs has been
found \citep{luhman00,briceno02,luhman03b}, although a more recent
determination of the star/brown dwarf fraction in this last region seems to indicate that it is similar to the Trapezium 
\citep{guieu06}. The study of the mass function in older clusters such as Praesepe \citep{gon06,boud10} 
and the Hyades \citep{bouvier08} also seems to indicate that there is a turnover 
in the hydrogen-burning mass limit, but in these cases the dynamical evolution of the clusters
 with age could explain this deficit. 
 In summary, for most of the very young star-forming regions and clusters, the mass spectrum is a 
rising function down to the substellar mass limit that can be represented by a potential law with an index in the range 0.4--1.2. 

\section{Conclusions}

We have presented a 1.12 deg$^2$ $IZ$ survey in the $\sigma$ Orionis 
cluster. From $I$, $I-Z$ color--magnitude diagrams , we have selected 153 objects in the magnitude 
range $I$=16--23\,mag; fainter candidates could be below the deuterium-burning mass limit. $J$-band data of candidates 
brighter than the completeness magnitude of 
the survey confirmed that 124 of them ($\sim$ 80\,\%) belong to the infrared photometric sequence of the cluster. 
The spatial distribution of these objects shows an increasing number of objects 
located at distances larger than 30 arcmin 
to the north and west of $\sigma$ Orionis that probably belong to a different population of the Orion's Belt, 
such as the $\zeta$ or $\epsilon$
Orionis clusters. The projected radial surface density of the 102 bona fide brown dwarf
 candidates in the cluster can be reproduced by a 
decreasing exponential law with a central density of 0.23 $\pm$ 0.03 obj/arcmin$^2$ and a 
characteristic radius of $\sim$ 1\,pc. The spatial distribution of more and less massive brown dwarfs is similar and hence, there is no 
clear mass segregation between these substellar objects.  
Comparison with Monte Carlo simulations shows no evidence of the presence of aggregation of brown dwarfs in the cluster. 
Based on near-infrared data from 2MASS and UKIDSS and mid-infrared data from IRAC/{\em Spitzer}, we conclude that 
$\sim$ 5--9\,\% and 30$\pm$7\,\%  of the brown dwarfs in the $\sigma$ Orionis cluster have $K$-band 
and mid-infrared excesses 
at wavelengths longer than 5.8\,$\mu$m, respectively, probably related to the presence of disks. 
The majority of them belong to the so called ``transition disks'', where a 
process of clearing out has occurred in their inner regions.  
We have estimated the substellar mass spectrum of the cluster, and found that this is a rising function towards
 lower masses 
and can be represented by a potential function ($dN/dm$ $\propto$ m$^{-\alpha}$) with a exponent $\alpha$ 
between 0.4 and 1.1 in the mass range 
between 0.11--0.013 M$_{\odot}$. 
These results are consistent with the majority of other studies of young open clusters and associations
and indicates that, although we cannot say that the substellar mass function is universal, its behavior is rather 
general.

\acknowledgments

We thank J. Licandro for his help in the acquisition of infrared data at the 
Carlos S\'anchez Telescope. We thank I. Baraffe and the Lyon group, 
F. D'Antona and A. Burrows for sending us electronic versions of their models. 
This work is based on observations obtained at: the Carlos  S\'anchez Telescope operated by 
the Instituto de Astrof\'{\i}sica de Canarias at the  Observatorio del Teide (Tenerife, Spain);
the Isaac Newton Telescope operated on the island of La Palma 
by the Isaac Newton Group in the Spanish Observatorio del Roque de los Muchachos of the 
Instituto de Astrof\'{\i}sica de Canarias; and
the German-Spanish Astronomical Center, Calar Alto, 
jointly operated by the Max-Planck-Institut f\"ur Astronomie Heidelberg and the 
Instituto de Astrof\'{\i}sica de Andaluc\'{\i}a (CSIC). 
The United Kingdom Infrared Telescope is operated by the Joint Astronomy Centre 
on behalf of the Science and Technology Facilities Council of the U.K. 
This publication makes use of data products from the Two Micron All Sky Survey, which is a joint project of 
the University of Massachusetts and the Infrared Processing and Analysis Center/California Institute of 
Technology, funded by the National Aeronautics and Space Administration and the National Science Foundation.
This research has made use of the VizieR catalogue access tool and the SIMBAD database, operated at CDS, 
Strasbourg, France, and IRAF, which is distributed by National Optical Astronomy
Observatories, which is operated by the Association of Universities for Research in Astronomy, Inc., under 
contract with the National Science Foundation. 
V. J. S. B. is partially supported by the Spanish Ram\'on y Cajal program. 
Partial financial support was provided by the Spanish Ministerio de Ciencia e Innovaci\'on projects AYA2007-67458, AYA2010-20535, AYA2010-21038-C03-01 and AYA2010-21038-C03-02. 
E. L. M. acknowledges support from Keck PI Data Analysis grant awarded by the Michelson Science Center.

{\it Facilities:} \facility{TCS (CAIN)}, \facility{INT (WFC)}, \facility{3.5\,m CAHA (Omega-Prime)}, \facility{2.2\,m CAHA (MAGIC)}.

\clearpage

\clearpage

\clearpage

\clearpage

\clearpage

\clearpage

\clearpage

\clearpage

\clearpage

\clearpage

\clearpage

\clearpage

\begin{deluxetable}{lcc}
\tabletypesize{\scriptsize}
\tablecaption{\label{tab1} Central coordinates of the four WFC pointings in present survey.}
\tablewidth{0pt}
\tablehead{
\colhead{Field} & \colhead{RA (J2000)} & \colhead{DEC (J2000)}  \\
      & \colhead{(h m s)}	 & \colhead{(o ' '')}
}

\startdata

SO1   & 5 39 06 & -02 16 17  \\
SO2   & 5 37 19 & -02 18 04 \\
SO3   & 5 36 57 & -02 50 01  \\
SO4   & 5 39 25 & -02 50 10  \\
\enddata
\end{deluxetable}

\clearpage

\LongTables
\begin{landscape}

\begin{deluxetable}{lccccccccccc}
\setlength{\tabcolsep}{0.03in} 
\tabletypesize{\scriptsize}
\tablecaption{\label{tab2} Photometry and coordinates of survey candiates}
\tablewidth{0pt}
\tablehead{
\colhead{Name} & \colhead{prev. ID.} & \colhead{$I$} & \colhead{$I-Z$} & \colhead{$J$} & \colhead{$I-J$} &  
\multicolumn{3}{c}{\underline{\qquad \qquad 2MASS photometry \qquad \qquad }} & \colhead{R.A.(J2000)}  & \colhead{DEC.(J2000)} & \colhead{Membership}\\  
\colhead{   }  &                     &               &                 &               &                 &  
\colhead{$J$} & \colhead{$H$} & \colhead{$K_{\rm s}$}                                   & \colhead{($^{h}$ m s)}& \colhead{($^{o}$ ' '')} & \colhead{}\\
}

\startdata

&&&&& SO1 field\\
\hline
\hline
&&&&& Detector 1\\
\hline

S\,Ori\,J053910.02--022811.5 &      (4)       &  16.080$\pm$0.040  &  0.660$\pm$0.040	&  14.550$\pm$0.050\tablenotemark{a}  &  1.530$\pm$0.060 &  14.600$\pm$0.030  &  14.000$\pm$0.040  &  13.780$\pm$0.050 & 05 39 10.02 & -02 28 11.5 & MC \\
S\,Ori\,J053944.33--023302.8 & S\,Ori\,11 (1) &  16.250$\pm$0.040  &  0.800$\pm$0.040	&  14.300$\pm$0.050\tablenotemark{a}  &  1.950$\pm$0.060 &  14.290$\pm$0.030  &  13.720$\pm$0.030  &  13.370$\pm$0.040 & 05 39 44.33 & -02 33 02.8 & MC \\
S\,Ori\,J053911.40--023332.7 &      (3)       &  16.330$\pm$0.040  &  0.720$\pm$0.040	&  14.450$\pm$0.100\tablenotemark{b}  &  1.880$\pm$0.110 &  14.450$\pm$0.030  &  13.930$\pm$0.030  &  13.570$\pm$0.040 & 05 39 11.40 & -02 33 32.7 & MC \\
S\,Ori\,J053848.10--022853.7 & S\,Ori\,15 (1) &  16.380$\pm$0.040  &  0.810$\pm$0.040	&  14.480$\pm$0.050\tablenotemark{a}  &  1.910$\pm$0.060 &  14.470$\pm$0.030  &  13.840$\pm$0.030  &  13.440$\pm$0.040 & 05 38 48.10 & -02 28 53.7 & MC \\
S\,Ori\,J053849.29--022357.6 &      (3)       &  16.410$\pm$0.040  &  0.670$\pm$0.040	&  14.230$\pm$0.060\tablenotemark{b}  &  2.180$\pm$0.070 &  14.360$\pm$0.030  &  13.700$\pm$0.030  &  13.200$\pm$0.030 & 05 38 49.29 & -02 23 57.6 & MC \\
S\,Ori\,J053907.60--022905.6 & S\,Ori\,20 (1) &  16.920$\pm$0.040  &  0.800$\pm$0.040	&  14.900$\pm$0.050\tablenotemark{a}  &  2.020$\pm$0.060 &  14.960$\pm$0.040  &  14.340$\pm$0.040  &  13.900$\pm$0.050 & 05 39 07.60 & -02 29 05.6 & MC \\
\enddata
\tablerefs{
(1)  \citealt{bejar99} (BZOR)
(2)  \citealt{osorio00} 
(3)  \citealt{bejar01} (BMZO)
(4)  \citealt{osorio02b}
(5)  \citealt{bejar04a} 
(6)  \citealt{kenyon05}
(7)  \citealt{burn05} 
(8)  \citealt{cab04,cab07a} 
(9)  \citealt{cab08a} 
(10) \citealt{lodieu09} }

\tablenotetext{a}{Photometry from Omega-Prime}
\tablenotetext{b}{Photometry from CAIN calibrated with 2MASS}
\tablenotetext{c}{Photometry from UKIDSS}
\tablenotetext{d}{Photometry from MAGIC}

\tablenotetext{MC}{Good cluster Member Candidate}
\tablenotetext{NM}{Probable Non Member}
\tablenotetext{OB}{Possibly belonging to the Orion Background population}

\end{deluxetable}

\begin{deluxetable}{lcccccccccccc}
\setlength{\tabcolsep}{0.02in} 
\tabletypesize{\scriptsize}
\tablecolumns{13}
\tablewidth{0pc}
\tablecaption{\label{tab3} UKIDSS and IRAC photometry}
\tablehead{
\colhead{ }    & \colhead{ } & \multicolumn{5}{c}{UKIDSS} & \colhead{ } & 
\multicolumn{4}{c}{IRAC/Spitzer}  & \colhead{ } \\
\cline{3-7} \cline{9-12} \\
\colhead{ID} & \colhead{$I$} &  
\colhead{$Z$} & \colhead{$Y$} & \colhead{$J$} & \colhead{$H$} & \colhead{$K$} & \colhead{} &
\colhead{[3.6]} & \colhead{[4.5]} & \colhead{[5.4]} & \colhead{[8.0]} & \colhead{Near/Mid} \\
\colhead{} & \colhead{(mag)} &  
\colhead{(mag)} & \colhead{(mag)} & \colhead{(mag)} & \colhead{(mag)} & \colhead{(mag)} & \colhead{} &
\colhead{(mag)} & \colhead{[(mag)} & \colhead{(mag)} & \colhead{(mag)} & \colhead{IR excess} }
\startdata
S\,Ori\,J053910.02--022811.5   &  16.080$\pm$0.040  &	15.504$\pm$0.005    &  15.044$\pm$0.004    &   14.533$\pm$0.004  &  13.990$\pm$0.004	&  13.706$\pm$0.004    &  &  13.37$\pm$0.04    &  13.51$\pm$0.04   &  13.41$\pm$0.06  &  13.25$\pm$0.11  &	 N/N	      \\
S\,Ori\,J053944.33--023302.8   &  16.250$\pm$0.040  &	15.581$\pm$0.005    &  14.902$\pm$0.004    &   14.282$\pm$0.004  &  13.788$\pm$0.003	&  13.433$\pm$0.003    &  &  13.24$\pm$0.04    &  13.18$\pm$0.05   &  12.99$\pm$0.04  &  12.87$\pm$0.08  &	 N/N	      \\
S\,Ori\,J053911.40--023332.7   &  16.330$\pm$0.040  &	15.687$\pm$0.006    &  15.057$\pm$0.004    &   14.437$\pm$0.004  &  13.904$\pm$0.004	&  13.581$\pm$0.003    &  &  13.30$\pm$0.04    &  13.20$\pm$0.03   &  13.10$\pm$0.04  &  13.02$\pm$0.09  &	 N/N	      \\
S\,Ori\,J053848.10--022853.7   &  16.380$\pm$0.040  &	15.650$\pm$0.006    &  14.991$\pm$0.004    &   14.392$\pm$0.004  &  13.862$\pm$0.004	&  13.488$\pm$0.003    &  &  13.08$\pm$0.05    &  12.75$\pm$0.06   &  12.38$\pm$0.03  &  11.77$\pm$0.05  &	 N/Y	      \\
S\,Ori\,J053849.29--022357.6   &  16.410$\pm$0.040  &	15.851$\pm$0.006    &  15.241$\pm$0.005    &   14.536$\pm$0.004  &  14.115$\pm$0.004	&  13.524$\pm$0.003    &  &  12.76$\pm$0.04    &  12.52$\pm$0.04   &  11.94$\pm$0.02  &  11.58$\pm$0.04  &	 Y/Y	      \\
S\,Ori\,J053907.60--022905.6   &  16.920$\pm$0.040  &	16.201$\pm$0.007    &  15.553$\pm$0.006    &   14.874$\pm$0.005  &  14.299$\pm$0.005	&  13.928$\pm$0.004    &  &  13.54$\pm$0.06    &  13.60$\pm$0.05   &  13.43$\pm$0.05  &  13.57$\pm$0.16  &	 N/N	      \\
S\,Ori\,J053926.77--022614.3   &  18.250$\pm$0.040  &	17.633$\pm$0.017    &  16.888$\pm$0.014    &   16.224$\pm$0.012  &  15.688$\pm$0.016	&  15.261$\pm$0.015    &  &	 \nodata       &  14.78$\pm$0.09   &  14.30$\pm$0.16  &  13.63$\pm$0.16  &	 N/\nodata    \\
\enddata
\end{deluxetable}

\clearpage
\end{landscape}

\begin{deluxetable}{lcccccccc}
\tabletypesize{\scriptsize}
\tablecaption{\label{tab4} Comparison of the $J$-band photometry of CAIN and Omega-Prime with 2MASS and UKIDSS.}
\tablewidth{0pt}
\tablehead{
\colhead{} & \colhead{Average} & \colhead{Error}  & \colhead{Std. dev.}	 & \colhead{Num.} & 
\colhead{Average\tablenotemark{a}} & \colhead{Error\tablenotemark{a}}  & \colhead{Std. dev.\tablenotemark{a}}	 & \colhead{Num.\tablenotemark{a}} \\
\colhead{} & \colhead{mag} & \colhead{mag}  & \colhead{mag}	 & \colhead{(obj.)} & 
\colhead{mag} & \colhead{mag}  & \colhead{mag}	 & \colhead{(obj.)} 
}

\startdata

 J(CAIN)--J(2MASS)    & 0.014 & 0.020 &  0.110 & 31  &  0.014 & 0.020 & 0.110 & 31 \\
 J(CAIN)--J(UKIDSS)   & 0.031 & 0.019 &  0.150 & 68  &  0.002 & 0.018 & 0.140 & 59 \\
 J(Omega)--J(2MASS)   & 0.002 & 0.015 &  0.090 & 35  & -0.014 & 0.011 & 0.060 & 33 \\
 J(Omega)--J(UKIDSS)  & 0.050 & 0.013 &  0.080 & 41  &  0.030 & 0.009 & 0.050 & 37 \\

\enddata
\tablenotetext{a}{Estimated quantities after discarding objects that deviate by more than three times their error bars.}
\end{deluxetable}

\begin{deluxetable}{lcccccc}
\tabletypesize{\scriptsize}
\tablecaption{\label{tab5} Estimated number of late M and L field dwarf contaminants.}
\tablewidth{0pt}
\tablehead{
\colhead{$\Delta$\,I} & \colhead{M5--M6} & \colhead{M6--M7}  & \colhead{M7--M9}	 & \colhead{M9--L0} & \colhead{L0--L4} & \colhead{All } \\
\colhead{(mag)}       & \colhead{}       & \colhead{}        & \colhead{}	 & \colhead{}       & \colhead{}       & \colhead{Sp. Type}   
}

\startdata

16--17  & 0.3     & 0.1     & 0.0 & 0.0 &  0.0 &  0.4 \\
17--18  & 1.3     & 0.4     & 0.0 & 0.0 &  0.0 &  1.7 \\
18--19  & 2.6     & 1.7     & 0.2 & 0.0 &  0.0 &  4.5 \\
19--20  & \nodata & 3.7     & 0.7 & 0.1 &  0.0 &  4.5 \\
20--21  & \nodata & \nodata & 2.7 & 0.5 &  0.2 &  3.4 \\
21--22  & \nodata & \nodata & 0.3 & 1.8 &  0.7 &  2.8 \\
All     & 4.2     & 5.9     & 3.9 & 2.4 &  0.9 & 17.3 \\
\enddata
\end{deluxetable}

\end{document}